\documentclass[aps,pra,superscriptaddress,twocolumn]{revtex4}
\usepackage{graphicx}
\usepackage{bm}
\usepackage{dsfont}

\usepackage{amsmath,amssymb}

\usepackage[hang,normalsize]{subfigure}
\usepackage{color,psfrag}
\usepackage{pstool}

\usepackage{array}
\newtheorem{theorem}{Theorem}
\newtheorem{lemma}{Lemma}
\newtheorem{definition}{Definition}

\usepackage{bbm} % use \mathbbm{1}

\newcommand{\id}{\mathbbm{1}}
\newcommand{\cc}{\mathbbm{C}}
\newcommand{\nn}{\mathbbm{N}}
\renewcommand{\vec}[1]{\text{\boldmath$#1$}}

\usepackage{color}

\usepackage{hyperref}

\begin{document}
\preprint{}
\title{Scalable Reconstruction of Density Matrices}

\author{T.\ Baumgratz}
\affiliation{Institut f\"{u}r Theoretische Physik, Albert-Einstein-Allee 11,
Universit\"{a}t Ulm, 89069 Ulm, Germany}
\affiliation{Center for Integrated Quantum Science and Technology, Universit\"{a}t Ulm, 89069 Ulm, Germany}

\author{D.\ Gross}
\affiliation{Physikalisches Institut, Hermann-Herder-Stra\ss e 3,
Albert-Ludwigs Universit\"{a}t Freiburg, 79104 Freiburg, Germany}

\author{M.\ Cramer}
\affiliation{Institut f\"{u}r Theoretische Physik, Albert-Einstein-Allee 11,
Universit\"{a}t Ulm, 89069 Ulm, Germany}
\affiliation{Center for Integrated Quantum Science and Technology, Universit\"{a}t Ulm, 89069 Ulm, Germany}

\author{M.B.\ Plenio}
\affiliation{Institut f\"{u}r Theoretische Physik, Albert-Einstein-Allee 11,
Universit\"{a}t Ulm, 89069 Ulm, Germany}
\affiliation{Center for Integrated Quantum Science and Technology, Universit\"{a}t Ulm, 89069 Ulm, Germany}

%\date{\today}

\begin{abstract}
Recent contributions in the field of quantum state tomography have shown that, despite the exponential growth of Hilbert space with the number of subsystems, tomography of one-dimensional quantum systems may still be performed efficiently by tailored reconstruction schemes. Here, we discuss a scalable method to reconstruct mixed states that are well approximated by matrix product operators. The reconstruction scheme only requires local information about the state, giving rise to a reconstruction technique that is scalable in the system size. 
It is based on a constructive proof that generic matrix product operators are fully determined by their local reductions. We discuss applications of this scheme for simulated data and experimental data obtained in an ion trap experiment. 

\end{abstract}

\maketitle

The complexity of many-body systems is one of the most intriguing, but
at the same time daunting, features of quantum mechanics. The {\em curse
of dimensionality}, namely the exponential growth 
of the descriptive complexity of even pure states,
is a property of
quantum mechanics which clearly distinguishes it from classical
physics. Therefore, in general, the number of variables required to
uniquely determine a quantum state increases in accordance with the
growth of the Hilbert space exponentially. 

Quantum state tomography addresses the problem of completely
characterizing a state of a physical system by measuring a complete
set of observables that determine the state uniquely \cite{paris04}.
As the complexity of quantum operations implemented in laboratories
steadily increases \cite{haeffner05,leibfried05,monz11,yao12}, the demand
for a reliable and scalable tomography 
\cite{cramer10, scalabletomo}
of prepared states is high and
of considerable importance for the future of quantum technologies. The
ability to store and manipulate interacting quantum many-body systems,
such as linearly arranged ions in an ion trap, enhanced rapidly during
the last years. Soon, if not already, the number of particles controllable
in such systems will cross the threshold for which conventional methods
of full quantum state tomography fail due to both the limited time that is
realistically available for the experiment and the limitations to the
resources that are available for the classical post-processing of the
experimental data \cite{haeffner05}. Further, while most experiments
have so far focused on the controlled creation of pure states and scalable
reconstruction methods have been tailored to the pure setting \cite{cramer10, scalabletomo},
experimental simulations of open system dynamics have begun to emerge \cite{open_system}, calling for efficient tomography of mixed states. 

The experimental time requirement is defined by the total number of measurements which have to be done to reconstruct the state faithfully; i.e., one has to consider the system size and the number of repetitions to obtain sufficient statistics \cite{james01}. The post-processing resources are determined by the individual tomography scheme and in particular by the representation of the state. Clearly, full quantum state tomography where the state is represented by an exponentially large number of variables will require an exponentially increasing computational power and is hence infeasible already for, e.g., trapped ion experiments available today \cite{monz11}. 
But many naturally occurring quantum states and many states of interest for quantum information tasks are completely characterized by a number of variables scaling moderately in the number of particles:
ground states of gapped local Hamiltonians \cite{hastings07,eisert10,plenio05}, thermal states of local Hamiltonians \cite{hastings06,eisert10}, the $W$ state, the GHZ state, and cluster states are all matrix product operators (matrix product states if they are pure) of low dimension, or very well approximated by them. These states are parametrized by a linear number of matrices of low bond-dimension. The key insight here is not that these states are matrix product operators or states ({\em any} state is a matrix product operator, respectively state) but that the matrix dimension is low, in particular independent of the system size. This solves one issue concerning the post-processing side of the problem mentioned above as these states may be stored efficiently on a classical computer. On the other hand, as we will see, generic matrix product operators
are not only completely determined by a linear number of local observables but may also be efficiently reconstructed from such local measurements, which makes the formalism we present here a powerful tool for quantum state tomography of mixed states \cite{footnote}. 

Recently, it has been demonstrated that the reconstruction of pure quantum states for large systems can be possible with the knowledge of local information only~\cite{cramer10}. The scheme presented in the latter reference relies on an efficient version of an iterative method first introduced in the context of matrix completion. While the original method comes with a convergence proof, this guarantee is lost in the efficient version. Here, we take a different approach, which works provably under a mild technical assumption on the state to be reconstructed and is not restricted to pure states. We present extensive numerical results for states not meeting the assumption guaranteeing uniqueness of the reconstructed density matrix.
In this manuscript we consider $N$ $d$-level subsystems aligned in a one-dimensional geometry, e.g., a chain of qubits ($d=2$). The aim is to reconstruct a mixed state $\hat{\varrho}$ from local information only. The local information we have in mind are estimates to all reductions of the state to a fixed number $R$ of contiguous sites. These may be obtained by estimating the expectation values of an informationally complete set of observables on the $R$ sites. Note that we do not require the estimates to the reductions to be states; i.e., empirical estimates to the expectation values of local observables suffice and post-processing such as maximum likelihood estimation is not required.
As $R$ is fixed, this corresponds to an experimental effort that is linear in the system size $N$. 
 We present a computationally cheap tomography scheme which scales polynomially in the system size $N$ and succeeds provably under a certain technical  {\em invertibility condition} on $\hat{\varrho}$. We demonstrate numerically and at the hand of experimental data (a W state on $8$ qubits created in an ion trap experiment~\cite{haeffner05}) that it still works well when this condition is not necessarily met. 

%{\em The scheme.---}
We begin our exposition of the tomography scheme by introducing some notation. We denote the to-be-reconstructed state by $\hat{\varrho}$. The input to the reconstruction scheme are estimates of expectation values which completely specify 
all reductions of $\hat{\varrho}$ to $R$ contiguous sites. We denote these reductions by $\hat{\varrho}_k$, $k=1,\dots, N-R+1$. Put mathematically, $\hat{\varrho}_k=\text{tr}_{\{1,\dots,k-1\}\cup\{k+R,\dots,N\}}[\hat{\varrho}]$, i.e., the trace over all but the $R$ sites $\{k,\dots,k+R-1\}$. Now let $\{\hat{P}_i^{(\alpha)}\}_{\alpha=1,\dots,d^{2}}$ be a complete operator basis for the site $i$. A common  choice for spin-$1/2$ particles is given by $\hat{P}_i^{(1)}=\id_i/\sqrt{2}$, $\hat{P}_i^{(2)}=\hat{\sigma}_i^x/\sqrt{2}$, $\hat{P}_i^{(3)}=\hat{\sigma}_i^y/\sqrt{2}$, $\hat{P}_i^{(4)}=\hat{\sigma}_i^z/\sqrt{2}$, i.e., the orthonormal Pauli spin basis.
We may then write
\begin{equation}
\hat{\varrho}_k=\sum_{\alpha_1,\dots,\alpha_R}\!\!\!\bigl\langle\hat{P}_k^{(\alpha_1)}\cdots \hat{P}_{k+R-1}^{(\alpha_R)}\bigr\rangle_{\hat{\varrho}}\,\hat{P}_k^{(\alpha_1)}\cdots \hat{P}_{k+R-1}^{(\alpha_R)},
\end{equation}
i.e., the $\hat{\varrho}_k$ are completely specified by the local expectation values $\langle\hat{P}_k^{(\alpha_1)}\cdots \hat{P}_{k+R-1}^{(\alpha_R)}\rangle_{\hat{\varrho}}=\text{tr}[\hat{\varrho}\hat{P}_k^{(\alpha_1)}\cdots \hat{P}_{k+R-1}^{(\alpha_R)}]$, estimates to which are the input to our tomography scheme.

\begin{figure}[t]
	\begin{center}
		\includegraphics[width=\columnwidth]{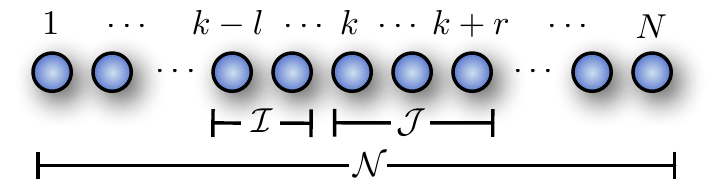}
	\end{center}
	\caption{Definition of the sets $\mathcal{N} = \{1,\dots,N\}$ and $\mathcal{I},\mathcal{J}\subset \mathcal{N}$. The linear map 
$E^{\mathcal{J}}_{\mathcal{I}}(\hat{X}) = \text{tr}_{\mathcal{J}}[\hat{X}\hat{O}_{\mathcal{I}\cup\mathcal{J}}]$ maps operators $\hat{X}$ (e.g., observables) living on set $\mathcal{J}$ into operators on set $\mathcal{I}$ by means of the reductions of $\hat{O}$ (e.g., the state) to $\mathcal{I}\cup\mathcal{J}$. }
\label{fig:ionchaindefinition}
\end{figure}

The $\hat{\varrho}_k$ completely specify the state $\hat{\varrho}$ if
a certain technical {\em invertibility
condition} is met. The proof is constructive and gives an explicit
method for obtaining $\hat{\varrho}$ from the $\hat{\varrho}_k$. It is partly inspired by the characterization of \emph{finitely
correlated states} (as opposed to $C^*$-finitely correlated states)
on infinite spin chains as described in the early
literature \cite[Prop.~2.1]{fannes92}. In addition to the fact that we are
working in a finite and non-translation invariant setting, the main
novel technical point here is that we only use \emph{local}
information, provided by the $\hat{\varrho}_k$.
To state the invertibility condition, we first
need to establish some notation.  We collect the $N$ sites of the one-dimensional
system in the set $\mathcal{N}=\{1,\dots,N\}$. For $\mathcal{I}\subset
\mathcal{N}$, we define the complex vector spaces $V_{\mathcal{I}}$
spanned by 
\begin{equation}
\Bigl\{\prod_{i\in \mathcal{I}}\hat{P}_i^{(\alpha_i)}\Bigr\}_{\alpha_i=1,\dots,d^{2}}.
\end{equation}
For given $\hat{O}\in V_{\mathcal{N}}$ and $\mathcal{I},\mathcal{J}\subset\mathcal{N}$ we define the linear map
$E_{\mathcal{I}}^{\mathcal{J}}:V_{\mathcal{J}}\rightarrow V_{\mathcal{N}\backslash\mathcal{I}}$ as
\begin{equation}
\hat{X}\mapsto E_{\mathcal{I}}^{\mathcal{J}}(\hat{X})=\text{tr}_{\mathcal{N}\backslash\mathcal{I}}[\hat{X}\hat{O}].
\end{equation}
We note that the map $E_{\mathcal{I}}^{\mathcal{J}}$ depends only on the reduction $\hat{O}_{\mathcal{I}\cup\mathcal{J}}=\text{tr}_{\mathcal{N}\backslash \mathcal{I}\cup\mathcal{J}}[\hat{O}]$ of $\hat{O}$ to sites $\mathcal{I}\cup\mathcal{J}$ as
\begin{equation}
E_{\mathcal{I}}^{\mathcal{J}}(\hat{X})=\text{tr}_{\mathcal{J}}[\hat{X}\hat{O}_{\mathcal{I}\cup\mathcal{J}}];
\end{equation}
this is illustrated in Fig. \ref{fig:ionchaindefinition}. Note that from now on we will only consider cases where $\mathcal{I}\cup\mathcal{J}$ is connected. 

\begin{definition}[Invertibility]
Let $l,r\in\nn$, $2\le l+r\le N-2$. If $\hat{O}$ is such that for all $k\in\nn$, $l\le k\le N-r-1$, the equality
\begin{equation}
\label{rank_eq}
\mathrm{rank}\Bigl[E_{\{k-l+1,\dots,k\}}^{\{k+1,\dots,k+r\}}\Bigr]=\mathrm{rank}\Bigl[E_{\{1,\dots,k\}}^{\{k+1,\dots,N\}}\Bigr]
\end{equation}
holds, we call $\hat{O}$ $(l,r)$-invertible.
\label{def:invertibilitycondition}
\end{definition}
We may now state the main theorem, a proof of which may be found  in Sec.~\ref{sec:ReconstructingInvertibleStates} of the Appendix.
\begin{theorem}Let $ l,r\in\nn$ such that $2\le l+r\le N-2$. Let $\hat{O}\in V_{\mathcal{N}}$ be $(l,r)$-invertible. Then, for all $\hat{X}_i\in V_{\{i\}}$, the equality
\begin{equation}
\label{last_step} 
\mathrm{tr}_{\mathcal{N}}[\hat{X}_1\cdots \hat{X}_{N}\hat{O}]=
\mathrm{tr}_{\mathcal{N}}[\hat{X}_1\cdots\hat{X}_{l}\hat{Y}_{l}\hat{O}]
\end{equation}
holds. Here, the $\hat{Y}_{l}\in V_{\{l+1,\dots,l+r\}}$ are recursively defined as follows. We set $\hat{Y}_{N-r}=\hat{X}_{N-r+1}\cdots\hat{X}_{N}$ and 
\begin{equation}
\label{equiv}
\hat{Y}_{k-1}=\bar{E}_{\{k-l,\dots,k-1\}}^{\{k,\dots,k+r-1\}}\Bigl(E_{\{k-l,\dots,k-1\}}^{\{k,\dots,k+r\}}(\hat{X}_{k}\hat{Y}_{k})\Bigr)
\end{equation}
for $k=l+1,\dots,N-r$. Here, the bar indicates the Moore-Penrose pseudoinverse.
\label{theorem:mainresult}
\end{theorem}
Note that, for Eq.\ (\ref{last_step}) the reduction of $\hat{O}$ to sites $\{1,\dots, l+r\}$ is needed; for the inverse we require the reduction of $\hat{O}$ to sites $\{k-l,\dots, k+r-1\}$, and for $E_{\{k-l,\dots,k-1\}}^{\{k,\dots,k+r\}}(\hat{X}_{k}\hat{Y}_{k})$ we require 
 the reduction of $\hat{O}$ to sites $\{k-l,\dots, k+r\}$. Hence, expectation values of the form
 $\text{tr}_{\mathcal{N}}[\hat{X}_1\cdots \hat{X}_{N}\hat{O}]$
are completely specified by reductions of $\hat{O}$ to the sites $\{k-l,\dots, k+r\}$, $k=l+1,\dots,N-r$, i.e., by all reductions to $R=r+l+1$ contiguous sites. By choosing the $\hat{X}_i$ to be the basis operators $\hat{P}_i^{\alpha_i}$, this implies that $(l,r)$-invertible operators $\hat{O}$ may be fully reconstructed from their reductions to $R$ consecutive sites, which is the same as knowing the expectation values
\begin{equation}
\text{tr}[\hat{P}_k^{\alpha_k}\cdots \hat{P}_{k+R-1}^{\alpha_{k+R-1}}\hat{O}],\;\;\; \alpha_i=1,\dots,d^{2},
\end{equation}
for all $k=1,\dots,N-R+1$.

One can prove that a vast majority of matrix product operators fulfil the invertibility condition; i.e., a vast majority of matrix product operators may be reconstructed from local reductions alone (see Sec.~\ref{sec:GenericMatrixProductOperatorsareInvertible} of the Appendix for a technical proof). As noted above, practically relevant states are (well approximated by) matrix product operators of low dimension; i.e., we expect the scheme to work for a large class of mixed states. Now, of course, experimentally, the exact expectation values even for states satisfying the invertibility condition are only known to within a certain statistical error (e.g., the estimated standard deviation of the mean after a finite number of measurements). This error propagates into the singular values of the map $E_{\{k-l,\dots,k-1\}}^{\{k,\dots,k+r-1\}}$. As this map needs to be inverted, even small errors on singular values close to zero will lead to a large error in the reconstruction. 
This issue may be avoided by using {\em stochastic robust approximation} techniques \cite{boyd04,lundeen09,zhang12} (see Sec.~\ref{sec:NonInvertibleInputs} of the Appendix for technical details). Before we apply the reconstruction scheme to experimental data, we present numerical results for states that do not necessarily fulfil the invertibility condition and for which the local expectation values are subject to inevitable statistical noise.

\begin{figure}[t]
	\begin{center}
		\includegraphics[width=0.9\columnwidth]{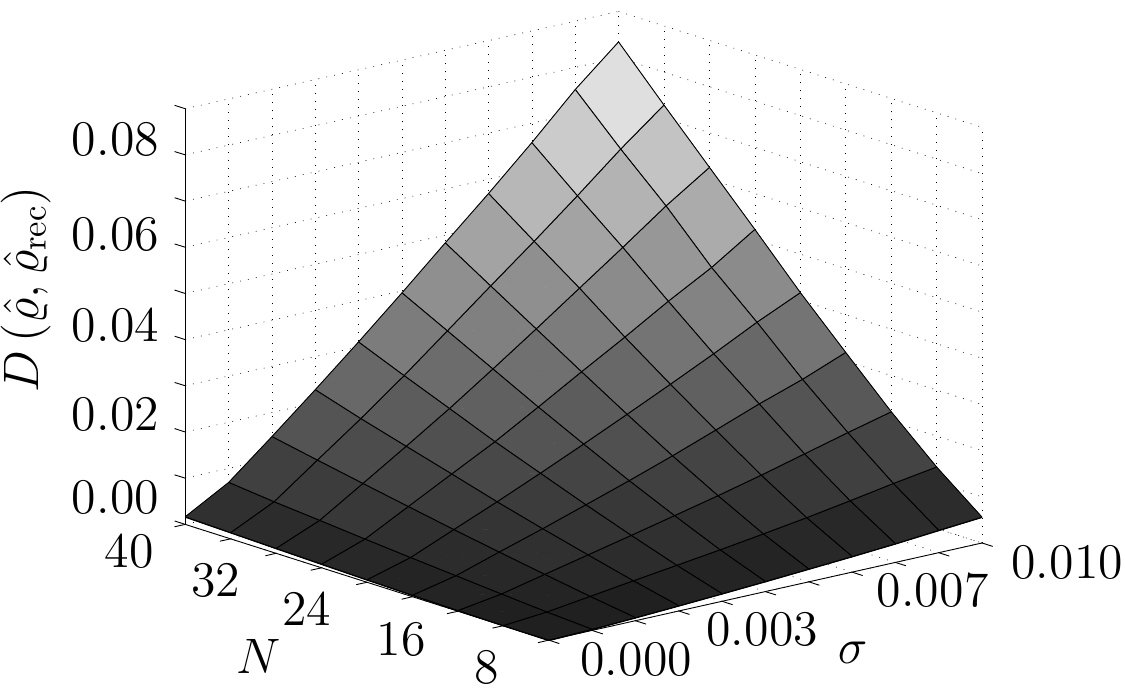}
	\end{center}
	\caption{Quality of our reconstruction scheme for thermal states of the Ising Hamiltonian in Eq.\ \eqref{eqn:isinghamiltonian} for $\beta=5$ and $R=5$, i.e., the state is reconstructed from local expectation values on five consecutive sites. For each pair $(N,\sigma)$, the plot shows the mean of the norm difference obtained from $100$ realizations and renormalized by the purity of the exact state, i.e. $D\left(\hat{\varrho},\hat{\varrho}_{\text{rec}}\right) = \| \hat{\varrho}_{\text{rec}}-\hat{\varrho}\|^{2} / \|\hat{\varrho} \|^{2}$. This corresponds to $100$ experiments, each of which carries  an uncertainty of $\sigma$ about the local expectation values.}
\label{fig:criticalIsing3d}
\end{figure}

%{\em Numerical Experiments.---}
We restrict our attention to qubits $d=2$, and illustrate the behaviour of the tomography scheme for thermal states of the Ising Hamiltonian at its quantum critical point
\begin{equation}
	\hat{H} = - \sum_{i=1}^{N-1} \hat{\sigma}_{i}^{x} \hat{\sigma}_{i+1}^{x} - 			
	\sum_{i=1}^{N} \hat{\sigma}_{i}^{z}.
\label{eqn:isinghamiltonian}
\end{equation}
We obtain the thermal states by an imaginary time evolution \cite{zwolak04,remark1} using the time evolving block decimation algorithm (TEBD). 
We simulate the measurements in the following way. We first compute the exact local expectation values $p_{\alpha_1,\dots,\alpha_R}^{k}=\langle\hat{\sigma}_k^{(\alpha_1)}\cdots \hat{\sigma}_{k+R-1}^{(\alpha_R)}\rangle_{\hat{\varrho}}$, $\alpha_i=0,x,y,z$, for all $k$.
Statistical noise is then simulated by adding random numbers (drawn from a Gaussian distribution with zero mean and standard deviation $\sigma$) to them. The resulting $\bar{p}_{\alpha_1,\dots,\alpha_R}^{k}$ then serve as the input to our reconstruction scheme. 
We compare the reconstructed state $\hat{\varrho}_{\text{rec}}$ to the exact state $\hat{\varrho}$ by computing  the Hilbert-Schmidt norm difference 
$D\left(\hat{\varrho},\hat{\varrho}_{\text{rec}}\right) = \| \hat{\varrho}_{\text{rec}}-\hat{\varrho}\|^{2} / \|\hat{\varrho} \|^{2}$.
To obtain meaningful results, we have rescaled the norm such that the deviations are measured in units of $\|\hat{\varrho}\|^2$, the natural length scale of the state to be learned. In Fig.\ \ref{fig:criticalIsing3d}, we show the norm difference for the exact and the reconstructed states as a function of the system size $N$ and the error $\sigma$. It indicates that, for given $N$, the error $D\left(\hat{\varrho},\hat{\varrho}_{\text{rec}}\right)$ scales roughly as $\sigma$; similarly, for given $\sigma$, it scales roughly as $N$. In Sec.~\ref{sec:NumericalExperimentsAppendix} of the Appendix we provide further numerical experiments analysing the performance of the algorithm for thermal states of random next-neighbour Hamiltonians and mixed states obtained by tracing out parts of a matrix product state in a larger Hilbert space. Again, these numerical results suggest that the scaling of our scheme is polynomial in both N and $\sigma$.

\begin{figure}[htb]
	\begin{center}
		\includegraphics[width=0.98\columnwidth]{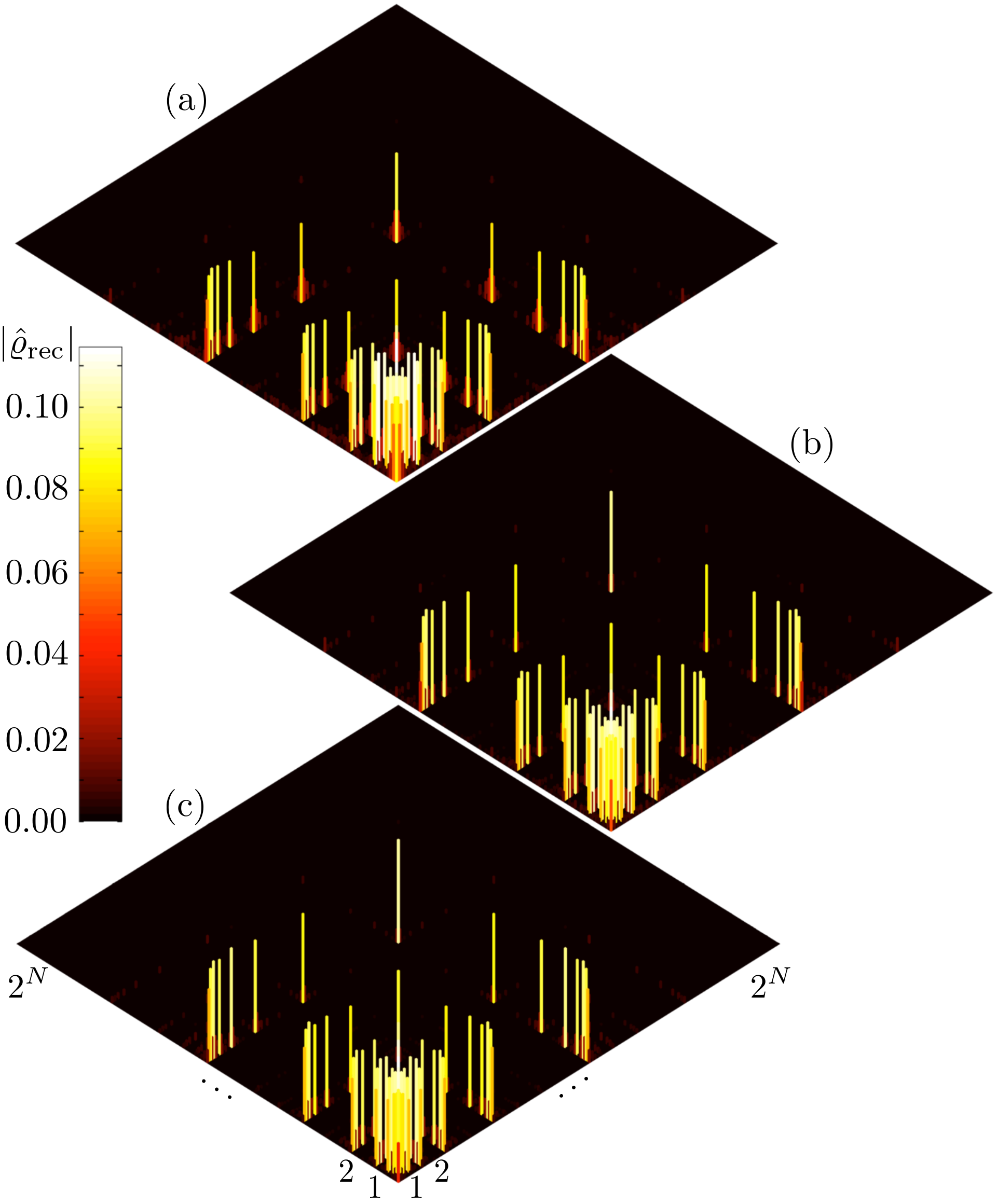}
	\end{center}
	\caption{Absolute value $|\hat{\varrho}_{\text{rec}}|$ of the corresponding reconstructed density matrix of the experimentally realized W state. (a): Reconstructed operator using the scheme described in this manuscript where the reductions to $R=3$ sites are known. (b): Estimate with $R=5$ sites. (c): Maximum likelihood estimate of full quantum state tomography (see also \cite{haeffner05}). The numbers $1,2,\ldots,2^N$ denote the entries of the density matrix $\hat{\varrho}_{\text{rec}}$. } 
\label{fig:ExperimentW8MPOReconstruction}
\end{figure}

Let us finally apply the reconstruction scheme to experimental data obtained in an ion trap experiment in a full quantum state tomography setting. The considered state is a W state implemented on $N=8$ qubits with local phases \cite{haeffner05}, i.e.,
\begin{equation}
\begin{split}
|W({\boldsymbol \phi})\rangle = &\left[ |0\ldots001\rangle + {\rm e}^{i\phi_{1}}| 0\ldots 010\rangle +  \right. \\
&\left. + \ldots + {\rm e}^{i\phi_{N-1}} |1\ldots 000\rangle \right] /\sqrt{N}.
\end{split}
\end{equation}
The available experimental data are the set of relative frequencies corresponding to $100$ measurements in each of the $3^N$ different basis rotations (measurements along the $X$, $Y$, and $Z$ directions). From these, we obtain maximum likelihood estimates to the reduced density matrices on all blocks of $R$ sites \cite{hradil04}. As described in the Appendix, we apply a {\em stochastic robust approximation} technique to avoid difficulties in ill-conditioned inversion problems making use of the Fisher information matrix of the local estimates \cite{hradil04}. Let us stress that the input to the reconstruction scheme are merely the relative frequencies corresponding to the measurements on all subsystems of $R$ contiguous sites and the total number of measurements. Absolute values of the reconstructed density matrices for $R=3$ and $R=5$ along with the maximum likelihood estimate obtained in the full tomography procedure \cite{haeffner05} are presented in Fig. \ref{fig:ExperimentW8MPOReconstruction}. Comparing the maximum likelihood estimate with our results we find for the renormalized Hilbert-Schmidt norm difference $D(\hat{\varrho}_{\text{ML}},\hat{\varrho}_{\text{rec}}) = 0.087$ for $R=3$ and 
$D(\hat{\varrho}_{\text{ML}},\hat{\varrho}_{\text{rec}}) = 0.012$ for $R=5$. For the full quantum state tomography experiment, maximizing the fidelity of the maximum likelihood estimate with respect to the local phases of a pure W state yields $f=\langle W({\boldsymbol \phi_{\text{opt}}})|\hat{\varrho}| W({\boldsymbol \phi_{\text{opt}}})\rangle = 0.722$ \cite{haeffner05}. With the matrix product operator scheme we achieve a fidelity of $f = 0.688$ for $R=3$ and $f = 0.718$ for $R=5$ with respect to the optimal W state $|W({\boldsymbol \phi_{\text{opt}}})\rangle$ revealing that the main contribution in our estimates stems from the same $|W({\boldsymbol \phi_{\text{opt}}})\rangle$ as in \cite{haeffner05}. We are only using local information and hence a local addressing of the ions in the trap is sufficient, resulting in the linear scaling of the scheme with the number of constituents. Further, the full maximum likelihood algorithm uses a huge amount of resources since it requires the storage and manipulation of $6^N$ measurement operators resulting in a time consuming post-processing. In contrast, our reconstruction takes about one second on a laptop given the local maximum likelihood estimates and the corresponding Fisher information matrices \cite{hradil04}. 

%{\em Conclusions.---}
In this work we have presented a scheme to reconstruct mixed states from local measurements efficiently. We have shown that, in principle, all states may be reconstructed from reductions to contiguous sets of sites alone and that the reconstruction is efficient with respect to the measurement time and the post-processing resources for practically relevant states. It should be noted, however, that our rigorous performance guarantees apply only when the model assumption of an essentially one-dimensional structure
is justified. As is the case for most statistical estimators, the scheme is
not suitable for \emph{model selection}; i.e.,\ it cannot certify
unconditionally from data alone that the model is valid. To investigate the latter issue, the impact of statistical noise and the performance of the reconstruction scheme for states that do not necessarily fulfil the condition which guarantees perfect reconstruction have been investigated for simulated states and experimental data in detail. 
For all simulations the Hilbert-Schmidt norm difference (normalized by the purity of the exact state) between the exact state and the reconstructed state was obtained and the numerical results  suggest that the quality of the reconstruction scales algebraically in $N$ and $\sigma$. The methods presented here hence pave the way for the reconstruction of mixed states of a large number of qubits. 

%\section{Acknowledgements}
We gratefully acknowledge H. H\"affner for providing experimental data, R. Rosenbach for 
results of the TEBD algorithm,  and K. Audenaert for fruitful discussions. Computations of the TEBD algorithm were performed on the bwGRiD \cite{BWGrid}. This
work was supported by the Alexander von Humboldt Foundation, the EU Integrated projects QESSENCE and SIQS, the BMBF
Verbundprojekt QuOReP, the
Excellence Initiative of the German Federal and State
Governments (grant ZUK 43), and the Swiss National Science Foundation.

%\newpage
%\setcounter{equation}{0}
\setcounter{theorem}{0}
%\renewcommand{\theequation}{A\arabic{equation}}
%\numberwithin{equation}{section}

\appendix
%\section{\label{sec:Appendix1}Implementation Details for the Maximum Likelihood
\section{\label{sec:ReconstructingInvertibleStates}Reconstructing Invertible States}
In the first section of the Appendix we provide a technical proof that expectation values of product observables with respect to all $(l,r)$-invertible states are fully determined by expectation values of observables acting only on a subset of the system. These observables can be determined recursively with the knowledge of all reduced density matrices to $R = r+l+1$ sites of the considered state. The proof of this lemma provides a scheme to directly determine a matrix product operator representation of the state. Before we proof the main result, let us recall the corresponding theorem in the main text, see theorem~\ref{theorem:mainresult}. 

\begin{theorem}Let $ l,r\in\nn$ such that $2\le l+r\le N-2$. Let $\hat{O}\in V_{\mathcal{N}}$ be $(l,r)$-invertible. Then, for all $\hat{X}_i\in V_{\{i\}}$, the equality
\begin{equation}
\label{last_step_appendix}
\mathrm{tr}_{\mathcal{N}}[\hat{X}_1\cdots \hat{X}_{N}\hat{O}]=
\mathrm{tr}_{\mathcal{N}}[\hat{X}_1\cdots\hat{X}_{l}\hat{Y}_{l}\hat{O}]
\end{equation}
holds. Here, the $\hat{Y}_{l}\in V_{\{l+1,\dots,l+r\}}$ are recursively defined as follows. We set $\hat{Y}_{N-r}=\hat{X}_{N-r+1}\cdots\hat{X}_{N}$ and 
\begin{equation}
\label{equiv}
\hat{Y}_{k-1}=\bar{E}_{\{k-l,\dots,k-1\}}^{\{k,\dots,k+r-1\}}\Bigl(E_{\{k-l,\dots,k-1\}}^{\{k,\dots,k+r\}}(\hat{X}_{k}\hat{Y}_{k})\Bigr)
\end{equation}
for $k=l+1,\dots,N-r$. Here, the bar indicates the Moore-Penrose pseudoinverse.
\end{theorem}

{\em Proof.} We start by showing that for all $k=l+1,\dots,N-r$, Eq.\ (\ref{equiv}) implies
\begin{equation}
\label{equiv1}
E_{\{1,\dots,k-1\}}^{\{k,\dots,k+r-1\}}(\hat{Y}_{k-1})=E_{\{1,\dots,k-1\}}^{\{k,\dots,k+r\}}(\hat{X}_{k}\hat{Y}_{k}).
\end{equation}
To this end, we define the linear map $\phi:\text{ran}[E_{\{1,\dots,k-1\}}^{\{k,\dots,k+r\}}]\rightarrow\text{ran}[E_{\{k-l,\dots,k-1\}}^{\{k,\dots,k+r\}}]$, where the domain and the range of $\phi$ are the ranges of the denoted linear maps, as
\begin{equation}
\begin{split}
\phi\bigl(E_{\{1,\dots,k-1\}}^{\{k,\dots,k+r\}}(\hat{Z})\bigr)&=\text{tr}_{1,\dots,k-l-1}\bigl[E_{\{1,\dots,k-1\}}^{\{k,\dots,k+r\}}(\hat{Z})\bigr]\\
&=E_{\{k-l,\dots,k-1\}}^{\{k,\dots,k+r\}}(\hat{Z}),
\end{split}
\end{equation}
i.e., $\text{ran}[\phi]=\text{ran}[E_{\{k-l,\dots,k-1\}}^{\{k,\dots,k+r\}}]$, and therefore, by the rank-nullity theorem,
\begin{equation}
\begin{split}
\text{dim}\bigl[\text{ker}[\phi]\bigr]&=\text{rank}[E_{\{1,\dots,k-1\}}^{\{k,\dots,k+r\}}]-\text{rank}[E_{\{k-l,\dots,k-1\}}^{\{k,\dots,k+r\}}]\\
&\le \text{rank}[E_{\{1,\dots,k-1\}}^{\{k,\dots,N\}}]-\text{rank}[E_{\{k-l,\dots,k-1\}}^{\{k,\dots,k+r-1\}}]\\
&=0
\end{split}
\end{equation}
due to the invertibility condition, see definition~\ref{def:invertibilitycondition} in the main text. Hence, $\phi(\hat{Z})=0$ is equivalent to $\hat{Z}=0$, i.e., Eq.\ (\ref{equiv1}) is equivalent to
\begin{equation}
\phi\bigl(E_{\{1,\dots,k-1\}}^{\{k,\dots,k+r\}}(\hat{Y}_{k-1}\otimes\id)\bigr)=\phi\bigl(E_{\{1,\dots,k-1\}}^{\{k,\dots,k+r\}}(\hat{X}_{k}\hat{Y}_{k})\bigr),
\end{equation}
which is implied by Eq.\ (\ref{equiv}). The theorem now follows by induction over $k=N-r-1,\dots,l+1$, starting at $k=N-r-1$:
the invertibility condition, see definition~\ref{def:invertibilitycondition} in the main text, guarantees the existence of $\hat{Y}_{N-r-1}\in V_{\{N-r,\dots,N-1\}}$ such that
\begin{equation}
\begin{split}
E^{\{N-r,\dots,N-1\}}_{\{1,\dots,N-r-1\}}(\hat{Y}_{N-r-1})=E^{\{N-r,\dots,N\}}_{\{1,\dots,N-r-1\}}(\hat{X}_{N-r}\hat{Y}_{N-r})\\
=E^{\{N-r,\dots,N\}}_{\{1,\dots,N-r-1\}}(\hat{X}_{N-r}\hat{X}_{N-r+1}\cdots \hat{X}_{N}),
\end{split}
\end{equation}
i.e., multiplying from the left by $\hat{X}_1\cdots\hat{X}_{N-r-1}$ and taking the trace over $\{1,\dots,N-r-1\}$, we find
\begin{equation}
\text{tr}_{\mathcal{N}}[
\hat{X}_1\cdots \hat{X}_{N}\hat{O}]=
\text{tr}_{\mathcal{N}}[
\hat{X}_1\cdots\hat{X}_{k}\hat{Y}_{k}\hat{O}]
\end{equation}
for $k=N-r-1$. Suppose now that this equality holds for $l<k\le N-r-1$ for some $\hat{Y}_k\in V_{\{k+1,\dots,k+r\}}$. We now show that it then also holds for $l\le k-1 \le N-r-2$. The invertibility condition, see definition~\ref{def:invertibilitycondition} in the main text, guarantees the existence of $\hat{Y}_{k-1}\in V_{\{k,\dots,k+r-1\}}$ such that
\begin{equation}
E^{\{k,\dots,k+r-1\}}_{\{1,\dots,k-1\}}(\hat{Y}_{k-1})=E^{\{k,\dots,k+r\}}_{\{1,\dots,k-1\}}(\hat{X}_{k}\hat{Y}_{k}),
\end{equation}
multiplying from the left by $\hat{X}_1\cdots\hat{X}_{k-1}$ and taking the trace over $\{1,\dots,k-1\}$, we find
\begin{equation}
\begin{split}
\text{tr}_{\mathcal{N}}[
\hat{X}_1\cdots \hat{X}_{N}\hat{O}]&=
\text{tr}_{\mathcal{N}}[
\hat{X}_1\cdots\hat{X}_{k}\hat{Y}_{k}\hat{O}]\\
&=
\text{tr}_{\mathcal{N}}[\hat{X}_1\cdots\hat{X}_{k-1}\hat{Y}_{k-1}\hat{O}],
\end{split}
\end{equation}
the desired equality for $l\le k-1 \le N-r-2$.

\section{\label{sec:GenericMatrixProductOperatorsareInvertible}Generic Matrix Product Operators are Invertible}
Here, we show that a vast majority of matrix product operators fulfil the invertibilty condition (see definition~\ref{def:invertibilitycondition} in the main text for details). Consider matrix product operators
\begin{equation}
\label{mpo}
\hat{O}=\sum_{\alpha_1,\dots,\alpha_N}P_1[\alpha_1]\cdots P_N[\alpha_N]\hat{P}_1^{(\alpha_1)}\cdots \hat{P}_N^{(\alpha_N)},
\end{equation}
with $P_1[\alpha]\in\cc^{1\times D_1}$, $P_N[\alpha]\in\cc^{D_N\times 1}$, and $P_i[\alpha]\in\cc^{D_i\times D_{i+1}}$ for $i=2,\dots,N-1$.
We assume w.l.o.g. that $\hat{P}_i^{(0)}\propto \id_i$ for all $i=1,\dots,N$.

\begin{lemma}
Let $ l,r\in\nn$ such that $2\le l+r\le N-2$. Let $\hat{O}$ be a matrix product operator as in Eq.\ (\ref{mpo}).
If
$\text{tr}[\hat{O}]\ne 0$ and for all $k\in\nn$, $l\le k\le N-r-1$, the sets
\begin{equation}
\{P_{k-l+1}[\alpha_{k-l+1}]\cdots P_{k}[\alpha_{k}]\}_{\alpha_{k-l+1},\dots,\alpha_{k}}
\end{equation}
 span $\cc^{D_{k-l+1}\times D_{k+1}}$ over $\cc$ and the sets
\begin{equation}
\{P_{k+1}[\alpha_{k+1}]\cdots
P_{k+r}[\alpha_{k+r}]\}_{\alpha_{k+1},\dots,\alpha_{k+r}}
\end{equation}
 span $\cc^{D_{k+1}\times D_{k+r+1}}$ over $\cc$, then $\hat{O}$ is $(l,r)$-invertible.
\end{lemma}
{\em Proof.} For $\hat{X}\in V_{\{k+1,\dots,k+r\}}$, 
\begin{equation}
\hat{X}=\sum_{\alpha_{k+1},\dots,\alpha_{k+r}}x_{\alpha_{k+1},\dots,\alpha_{k+r}}\hat{P}_{k+1}^{\alpha_{k+1}}\cdots
\hat{P}_{k+r}^{\alpha_{k+r}},
\end{equation}
we find
\begin{equation}
\begin{split}
&E_{\{k-l+1,\dots,k\}}^{\{k+1,\dots,k+r\}}(\hat{X})\\
&\hspace{0.3cm}\propto
\sum_{\alpha_{k-l+1},\dots, \alpha_{k}}\!\!\!\!\!\!\!\!\!P_1[1]\cdots P_{k-l}[1]
P_{k-l+1}[\alpha_{k-l+1}]\cdots P_{k}[\alpha_{k}]\\
&\hspace{0.5cm}\times
\sum_{\alpha_{k+1},\dots,\alpha_{k+r}}\!\!\!\!\!\!\!\!\!x_{\alpha_{k+1},\dots,\alpha_{k+r}}
P_{k+1}[\alpha_{k+1}]\cdots
P_{k+r}[\alpha_{k+r}]\\
&\hspace{2.5cm}\times
P_{k+r+1}[1]\cdots
 P_M[1]
\hat{ P}_{k-l+1}^{\alpha_{k-l+1}}\cdots \hat{P}_{k}^{\alpha_{k}}\\
&\hspace{0.3cm}=:\sum_{\alpha_{k-l+1},\dots, \alpha_{k}}\!\!\!\!\!\vec{w}^\dagger
P_{k-l+1}[\alpha_{k-l+1}]\cdots P_{k}[\alpha_{k}] X\vec{v}\\
&\hspace{3cm}\times\hat{ P}_{k-l+1}^{\alpha_{k-l+1}}\cdots \hat{P}_{k}^{\alpha_{k}}\\
&\hspace{0.3cm}=:\Gamma(X),
\end{split}
\end{equation}
where the matrix
\begin{equation}
\begin{split}
X&=\sum_{\alpha_{k+1},\dots,\alpha_{k+r}}\!\!\!\!\!\!\!\!\!x_{\alpha_{k+1},\dots,\alpha_{k+r}}
P_{k+1}[\alpha_{k+1}]\cdots
P_{k+r}[\alpha_{k+r}]\\
&\in\cc^{D_{k+1}\times D_{k+r+1}},
\end{split}
\end{equation}
the vectors
\begin{equation}
\begin{split}
\vec{v}&=
P_{k+r+1}[1]\cdots P_{M}[1]\in\cc^{D_{k+r+1}\times 1},\\
\vec{w}^\dagger&=P_1[1]\cdots P_{k-l}[1]\in\cc^{1\times D_{k-l+1}},
\end{split}
\end{equation}
and the mapping $\Gamma:\cc^{D_{k+1}\times D_{k+r+1}}\rightarrow V_{\{k-l+1,\dots,k\}}$.
Now, $\Gamma(X)=0$ is equivalent to
\begin{equation}
\begin{split}
0&=\vec{w}^\dagger
P_{k-l+1}[\alpha_{k-l+1}]\cdots P_{k}[\alpha_{k}] X\vec{v}\\
&=\text{tr}[P_{k-l+1}[\alpha_{k-l+1}]\cdots P_{k}[\alpha_{k}]X\vec{v}\vec{w}^\dagger]
\end{split}
\end{equation}
for all $\alpha_{k-l+1},\dots,\alpha_{k}$. Hence, if $\{P_{k-l+1}[\alpha_{k-l+1}]\cdots P_{k}[\alpha_{k}]\}_{\alpha_{k-l+1},\dots,\alpha_{k}}$ spans $\cc^{D_{k-l+1}\times D_{k+1}}$ over $\cc$, this is equivalent to $X\vec{v}\vec{w}^\dagger=0$.
Now, as $\vec{w}\ne \vec{0}$ (implied by $\text{tr}[\hat{O}]\ne 0$), this is equivalent to
$X\vec{v}=0$. Hence,
\begin{equation}
\text{ker}[\Gamma]=\left\{X\in\cc^{D_{k+1}\times D_{k+r+1}}\,\big|\,X\vec{v}=0\right\},
\end{equation}
i.e., the rank of $\Gamma$ is equal to
\begin{equation}
\begin{split}
D_{k+1}D_{k+r+1}-\text{dim}\left\{X\in\cc^{D_{k+1}\times D_{k+r+1}}\,\big|\,X\vec{v}=0\right\}.
\end{split}
\end{equation}
Now, if $\{P_{k+1}[\alpha_{k+1}]\cdots
P_{k+r}[\alpha_{k+r}]\}_{\alpha_{k+1},\dots,\alpha_{k+r}}$ spans $\cc^{D_{k+1}\times D_{k+r+1}}$ over $\cc$, we have
\begin{equation}
\text{ran}[E_{\{k-l+1,\dots,k\}}^{\{k+1,\dots,k+r\}}]=
\text{ran}[\Gamma],
\end{equation}
i.e., the rank of $E_{\{k-l+1,\dots,k\}}^{\{k+1,\dots,k+r\}}$ is equal to
\begin{equation}
\begin{split}
D_{k+1}D_{k+r+1}-\text{dim}\left\{X\in\cc^{D_{k+1}\times D_{k+r+1}}\,\big|\,X\vec{v}=0\right\}.
\end{split}
\end{equation}
As $\vec{v}\ne \vec{0}$ (implied by $\text{tr}[\hat{O}]\ne 0$), we may set $\vec{v}_1=\vec{v}$ and assume that there are vectors $\vec{v}_i\in \cc^{D_{k+r+1}\times 1}$, $i=2,\dots,D_{k+r+1}$, such that $\{\vec{v}_i\}_{i=1,\dots,D_{k+r+1}}$ is an orthogonal basis for $\cc^{D_{k+r+1}\times 1}$. Letting $\{\vec{u}_i\}_{i=1,\dots,D_{k+1}}$ an orthogonal basis for $\cc^{D_{k+1}\times 1}$, we may write
\begin{equation}
X=\sum_{i=1}^{D_{k+1}}\sum_{j=1}^{D_{k+r+1}}x_{i,j}\vec{u}_i\vec{v}_j^\dagger,
\end{equation}
i.e., $0=X\vec{v}=X\vec{v}_1$ is equivalent to $0=x_{i,1}$ for all $i=1,\dots,D_{k+1}$. Hence, the rank of $E_{\{k-l+1,\dots,k\}}^{\{k+1,\dots,k+r\}}$ is equal to
\begin{equation}
\begin{split}
D_{k+1}D_{k+r+1}-D_{k+1}(D_{k+r+1}-1)=D_{k+1}.
\end{split}
\end{equation}
Finally, 
\begin{equation}
\begin{split}
\text{rank}[
E_{\{1,\dots,k\}}^{\{k+1,\dots,N\}}] \le D_{k+1}.
\end{split}
\end{equation}

\section{\label{sec:NonInvertibleInputs}Non-invertible Inputs}
The main issue arising when applying the reconstruction scheme to experimental data is that the local reductions are not known exactly. But of course, we may simply use their estimates (e.g. direct inversions of the measurements, maximum likelihood estimates) as an input to compute the maps $E_{\{k-l,\dots,k-1\}}^{\{k,\dots,k+r-1\}}$ and $E_{\{k-l,\dots,k-1\}}^{\{k,\dots,k+r\}}$. However, as we need to compute the inverse of the former map, already a small uncertainty will lead to a large error in the inverse. This issue can be dealt with the method of {\em stochastic robust approximation} \cite{boyd04}. Before we introduce this regularization technique, let us first ease notation a bit. 
We write
\begin{equation}
\begin{split}
\hat{P}_i&=
\hat{P}_{k-l}^{(\alpha_{k-l})}\cdots \hat{P}_{k-1}^{(\alpha_{k-1})},\;\;\; i=1,\dots,d^{2l},\\
\hat{Q}_i&=
\hat{P}_k^{(\alpha_k)}\cdots \hat{P}_{k+r-1}^{(\alpha_{k+r-1})},\;\;\; i=1,\dots,d^{2r}.
\end{split}
\end{equation}
Suppose now that one had access to the exact local expectation values. The matrix representation, $A$, of $E_{\{k-l,\dots,k-1\}}^{\{k,\dots,k+r-1\}}$ would then be given by
\begin{equation}
\begin{split}
A_{i,j}&=\text{tr}_{k-l,\dots,k-1}[\hat{P}_i E_{\{k-l,\dots,k-1\}}^{\{k,\dots,k+r-1\}}(\hat{Q}_j)]\\
&=\text{tr}[\hat{P}_i \hat{Q}_j\hat{O}].
\end{split}
\end{equation}
Instead, we have only access to the noisy version of the entries $\text{tr}[\hat{P}_i \hat{Q}_j\hat{O}]$. Let us denote the resulting matrix by $B$. The errors in the measurements propagate into errors of the matrix $A$. In particular, we write $B = A + G$ where the matrix $G$ contains the errors due to imperfect measurements. Now, when applying the reconstruction scheme we have to solve linear equations of the form $B\vec{x} = \vec{e}$ where $B$ is as described above and $\vec{e}$ is known. Instead of directly inverting this equation, we take possible variations in the matrix $B$ into account, implying that the entries of $B$ are themselves prone to noise and attempting to undo the imperfect measurements. This is done by introducing the matrix $G^{\prime}$ and solving the statistical least-squares problem \cite{boyd04}
\begin{equation}
\text{argmin}\;\mathbbm{E}\left[ \| (B + G^{\prime}) \vec{x} - \vec{e} \|^{2} \right]
\label{eqn:statisticalleastsquareproblem}
\end{equation}
where $\mathbbm{E}$ denotes the expectation value and we try to annul the errors in $B$ by the random matrix $G^{\prime}$ which we assume to be a multivariate Gaussian distributed random matrix with zero mean and covariance matrix $\mathcal{C}^{\prime}$. This minimization problem can be rewritten as \cite{boyd04}
\begin{equation}
\text{argmin}\;\left[ \| B\vec{x} - \vec{e} \|^{2} + \vec{x}^{\mathrm T}P\vec{x} \right]
\end{equation}
with $P = \mathbbm{E}[ (G^{\prime})^{\mathrm T}G^{\prime}] $. This sort of regularization problems can be solved analytically. The solution is given by 
\begin{equation}
\vec{x} = \left( B^{\mathrm T}B + P \right)^{-1}  B^{\mathrm T} \vec{e}
\label{eqn:solutionstatisticalleastsquareproblem}
\end{equation}
where the entries of $P$ are closely related to the covariance matrix $\mathcal{C}^{\prime}$ of $G^{\prime}$
\begin{equation}
P_{k,l} = \sum_{i}\mathbbm{E}[ G^{\prime}_{i,k}G^{\prime}_{i,l}] = \sum_{i} \mathcal{C}^{\prime}_{(i,k),(i,l)}.
\end{equation}

It remains to find an appropriate model for the covariance matrix $\mathcal{C}^{\prime}$ of the assumed error $G^{\prime}$. In the numerical experiments we simulate statistical noise by adding independent random numbers (drawn from a Gaussian distribution with zero mean and standard deviation $\sigma$) to the expectation values of (unnormalized) Pauli strings. With this, the covariance matrix is proportional to the identity and taking the normalization factor into account we find $\mathcal{C}^{\prime}_{k,l} = \delta_{k,l} \cdot \sigma / \sqrt{d^{r+l}}$. Hence $P_{k,l} = \delta_{k,l}\cdot\sigma^{2}$, and the minimization problem reads 
\begin{equation}
\text{argmin}\;\left[ \| B\vec{x} - \vec{e} \|^{2} + \sigma^{2}\|\vec{x} \|^{2} \right]. 
\end{equation}
Problems of this form are known as Tikhonov regularizations \cite{boyd04,lundeen09,zhang12} with solution 
\begin{equation}
\vec{x} = \left( B^{\mathrm T}B + \sigma^{2} \right)^{-1}  B^{\mathrm T} \vec{e}. 
\end{equation}
Let us denote the singular values of $B$ by $s_1\ge\cdots \ge s_{\min\{d^{2l},d^{2r}\}}$ such that $B = U S V^{\mathrm T}$. Then, the solution of the statistical least-square problem \eqref{eqn:statisticalleastsquareproblem} in this scenario is given by $\vec{x} = \bar{B} \vec{e}$ with $\bar{B} = V \bar{S} U^{\mathrm T}$ where $\bar{S}$ is a diagonal matrix with entries $f_{i} / s_{i}$ and where $f_{i} = s_{i}^{2} / (s_{i}^{2} + \sigma^{2})$ is a smoothing factor suppressing the effect of the smallest singular values of $E^{k,\ldots,k+r-1}_{k-l,\ldots,k-1}$ in its inverse.

For the real experimental data we perform maximum likelihood locally to obtain estimates of the reduced density matrices. The remaining problem is to find an error model of the expansion coefficients of the maximum likelihood estimate in the Pauli basis, i.e., the entries of the matrix $B$. These coefficients can be modelled as the parameters which have to be estimated by the maximum likelihood scheme. To obtain an estimate of the error of these real parameters, we compute the Fisher information matrix $\mathcal{F}$ \cite{hradil04} for each subset whose inverse gives a lower bound on the covariance matrix of the matrix $B$ (with respect to the positive semidefinite cone). This is known as the Cram\'{e}r-Rao lower bound~\cite{hradil04}. Note that maximum likelihood estimates saturate this inequality asymptotically for a large number of measurements \cite{hradil04}. Writing $B = A + G$ where $G$ is a random matrix with zero mean the covariance matrix of B is equivalent to the covariance matrix $\mathcal{C}$ of $G$ and hence 
\begin{equation}
\mathcal{C} \geq \mathcal{F}^{-1}
\end{equation}
where 
\begin{equation}
\mathcal{F}_{(i,k),(j,l)} = \mathbbm{E}\left[ \frac{\partial \, \text{log}\,\mathcal{L}}{\partial \, B_{i,k}} \; \frac{\partial \, \text{log} \, \mathcal{L}}{\partial \, B_{j,l}} \right]
\end{equation}
is the Fisher information matrix with $\mathcal{L}$ the likelihood function and where $B_{i,k}$ denotes one entry of the matrix $B$, i.e., an expectation value of the maximum likelihood estimate with a normalized Pauli spin basis element. With this, we model the covariance matrix $\mathcal{C}^{\prime}$ of the random matrix $G^{\prime}$ in \eqref{eqn:statisticalleastsquareproblem} with the inverse of the Fisher information matrix. Hence, the matrices $P$ can be computed for all subsystems and the solution of the local inversion problems are given by Eq.~\eqref{eqn:solutionstatisticalleastsquareproblem}. Finally, let us stress that with this procedure the input to the reconstruction scheme are solely the relative frequencies of locally complete measurements obtained in the laboratory and the total number of performed measurements. 

\section{\label{sec:NumericalExperimentsAppendix}Numerical Experiments}

\begin{figure*}[tb]
	\begin{center}
		\includegraphics[width=0.95\textwidth]{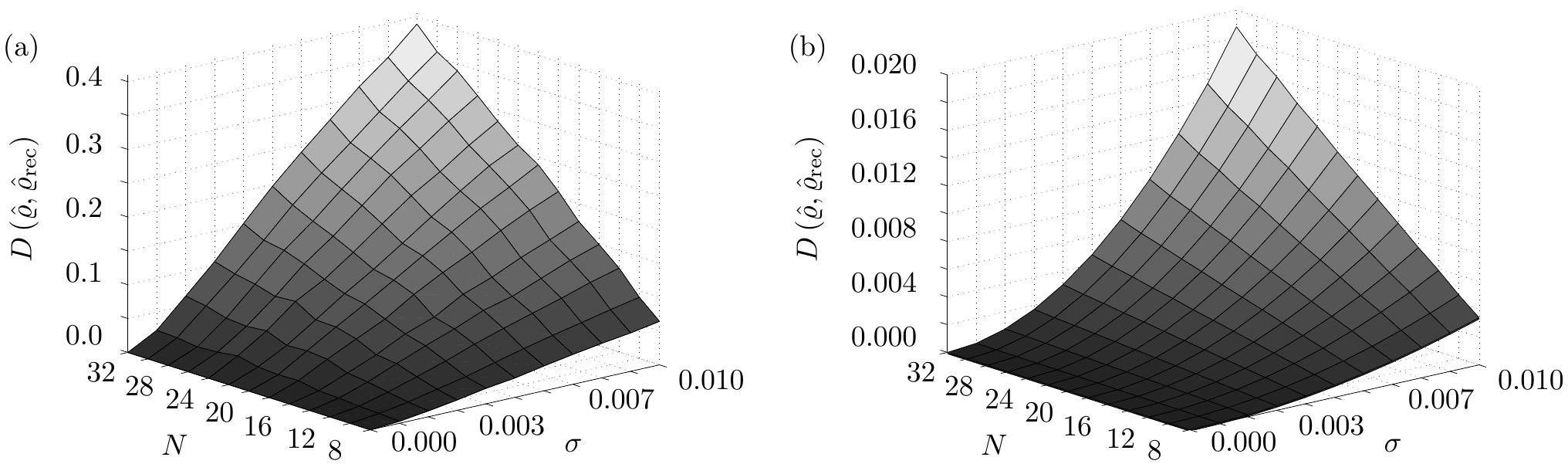}
	\end{center}
	\caption{Reconstruction errors for randomly chosen matrix product operators as described in the text with  $|\mathcal{N}_{\text{aux}}|=N$. The interaction is weak in a sense that we choose $t$ such that $t \|\hat{H}_{k}\|_{\text{op}} = 1/100$ for all $k$. For each pair $(N,\sigma)$ we draw $4000$ random states, simulate one measurement each and reconstruct the state with the disturbed local expectation values. The plot shows the mean values of the renormalized norm differences in dependence on the system size $N$ and the error in the measurements $\sigma$. (a) The states are reconstructed with $R=3$, i.e. measurements are done on all blocks of three contiguous sites. Here, for given $N$ ($\sigma$), the scaling of $D\left(\hat{\varrho},\hat{\varrho}_{\text{rec}}\right)$ is roughly linear in $\sigma$ ($N$). (b) Reconstruction with $R=5$. $D\left(\hat{\varrho},\hat{\varrho}_{\text{rec}}\right)$ improves significantly when measuring on larger blocks. } 
\label{fig:randominteractionMPO}
\end{figure*}

\begin{figure}[b]
	\begin{center}
		\includegraphics[width=0.92\columnwidth]{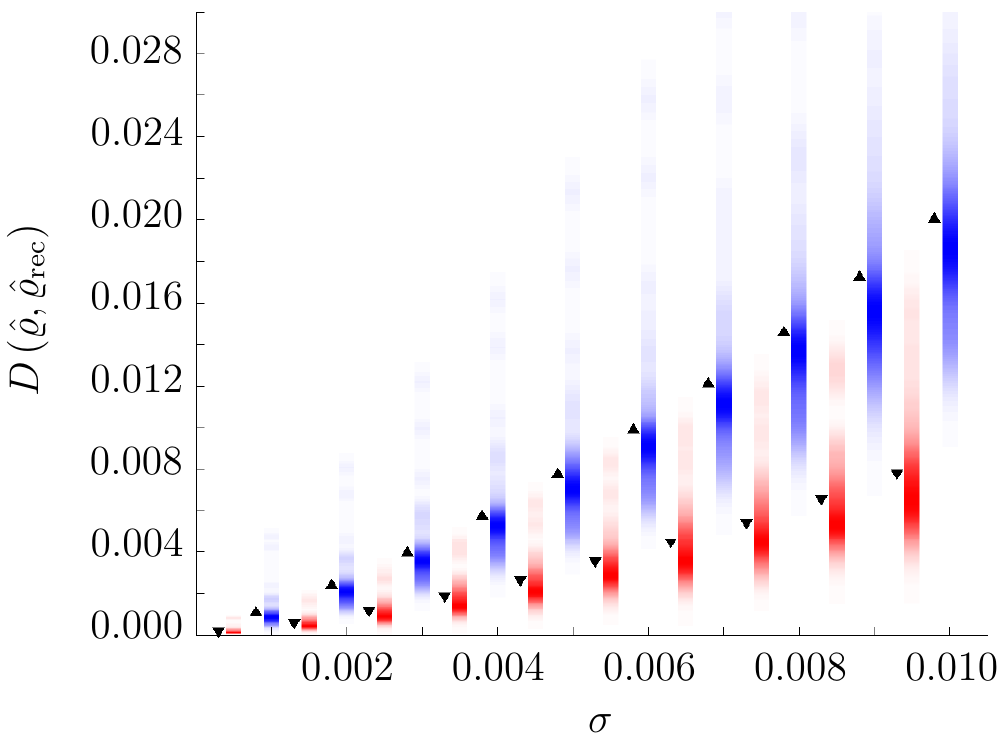}
	\end{center}
	\caption{Quality of our reconstruction scheme for thermal states of randomly chosen next-neighbour Hamiltonians as in Eq.\ \eqref{eqn:randomnextneighbourhamiltonian} with $\beta=2$ and $R=5$, i.e. the state is reconstructed from local expectation values on five consecutive sites. Downward-pointing triangles: system size $N=16$, upward-pointing triangles: system size $N=32$. We generate $50$ different random Hamiltonians and compute their corresponding thermal states using the TEBD algorithm. 
	For each state and pair $(N,\sigma)$, the density plot shows the simulations of one experiment carrying  an uncertainty of $\sigma$ about the local expectation values.  Mean values are indicated as triangles. } 
\label{fig:randomNextNeighbourHO}
\end{figure} 

In this section we continue the numerical analysis of the proposed algorithm for simulated states on large system sizes. In the main text we discussed the behaviour of the reconstruction scheme for thermal states of the Ising Hamiltonian at its quantum critical point. As a second numerical experiment let us analyse the behaviour of the algorithm for states which are exactly representable as matrix product operators satisfying the invertibility condition but subject to statistical noise. We pick such matrix product operators at random by generating a matrix product state with bond-dimension $D=d$ where the entries of the matrices defining the states are drawn from a Gaussian distribution with zero mean and standard deviation one. Then, we let these sites interact with an auxiliary system each of dimension $d$ according to the unitary $\hat{U}_{k} = {\rm e}^{-i\hat{H}_{k}t}$ for $k=1,\ldots,|\mathcal{N}_{\text{aux}}|$, where $\hat{H}_{k}$ is a two-particle interaction Hamiltonian acting on site $k$ and its auxiliary system with entries picked from a Gaussian distribution with zero mean and standard deviation one. Finally, we trace over the $|\mathcal{N}_{\text{aux}}|$ auxiliary sites to obtain a matrix product operator with bond-dimension $D=d^2$. From these states, we compute the exact local expectation values $p_{\alpha_{1},\ldots,\alpha_{R}}^{k}$, $\alpha_{i} = 0,x,y,z$ for all $k$, simulate the measurements by adding random numbers (drawn from a Gaussian distribution with zero mean and standard deviation $\sigma$), and reconstruct the state by means of the noisy local expectation values. Fig.~\ref{fig:randominteractionMPO} shows the results for different system sizes and different noise levels. Note that the bond-dimension of the estimate is fixed by the number of sites on which measurements are performed, see theorem~\ref{theorem:mainresult}. The larger these blocks (i.e., the larger $R$), the larger the bond-dimension of the estimate. This close connection between bond-dimension and block size can be seen in Fig. \ref{fig:randominteractionMPO}: Increasing the block size dramatically increases the accuracy of the estimate, suggesting the experimental strategy: The block size should be increased until a desired accuracy is reached or measurement time runs out, whichever happens first. Again, the numerical results suggest that the scaling of our scheme is polynomial in both, $N$ and $\sigma$. 

Thermal states of random next-neighbour Hamiltonians of the form 
\begin{equation}
	\hat{H}=\sum_{i=1}^{N-1} \hat{r}^{i}_{i,i+1}
\label{eqn:randomnextneighbourhamiltonian}
\end{equation}
serve as our last example. Here, the $\hat{r}^{i}_{i,i+1}$ are Hermitian matrices acting on sites $i$ and $i+1$ with entries that have real and imaginary part picked from a Gaussian distribution with zero mean and standard deviation one. Again, we use the TEBD \cite{zwolak04,remark1} algorithm to obtain the exact thermal states. 
%From these states, we compute the exact local expectation values $p_{\alpha_{1},\ldots,\alpha_{R}}^{k}$, $\alpha_{i} = 0,x,y,z$ for all $k$, simulate the measurements by adding random numbers (drawn from a Gaussian distribution with zero mean and standard deviation $\sigma$), and reconstruct the state by means of the noisy local expectation values. 
For each system size we generate $50$ random Hamiltonians and their corresponding thermal states and simulate one experiment for each $\sigma$ and state. Fig. \ref{fig:randomNextNeighbourHO} shows the error of the reconstructions as a function of the error of the measurements for two different system sizes. The densities illustrate the distribution of the error for the 50 different states while the black arrows indicate the mean. 

%\newpage


\begin{thebibliography}{00}
\bibitem{remark1}
To generate the thermal states we use a fourth order Trotter expansion and adjust the step size in the imaginary time evolution to $\delta T=10^{-3}$ for the Ising Hamiltonian and the random next-neighbour Hamitonians, respectively, i.e., $\hat{\varrho}=({\rm e}^{-\delta T  \hat{H}})^{\beta/{\delta T}}/Z$. Further, we keep the $100$ largest singular values in each decomposition, i.e., approximate the state as a matrix product operator with bond-dimension $D=100$.
\end{thebibliography}
\end{document}